\documentclass[a4paper,11pt]{article}
\usepackage{pos}

\def\MADMAX       {\mbox{{\sc Madmax}}} 

\title{First searches for axion and dark photon dark matter with MADMAX}

\manuallySeparateAuthors
\author*[a]{Fabrice Hubaut}
\author{ on behalf of the MADMAX collaboration}
\affiliation[a]{Aix Marseille Univ, CNRS/IN2P3, CPPM, Marseille, France}
\emailAdd{hubaut@in2p3.fr}

\abstract{The MAgnetized Disk and Mirror Axion eXperiment (\MADMAX{}) is a future experiment aiming to detect
dark matter axions from the galactic halo by resonant conversion to photons in a strong magnetic field.
It uses a novel concept based on a stack of dielectric disks in front of a mirror, called booster, to 
enhance the potential signal from axion-photon conversion over a significant mass range. 
In its final version, \MADMAX{} aims to scan the uncharted QCD axion mass range around 100 $\mu$eV, 
favored by post-inflationary theories.
Several small scale prototype systems have been tested these last three years, allowing to validate the 
dielectric haloscope concept and perform competitive axion and dark photon dark matter searches. 
This contribution presents the current status of the experiment and its prototypes, including the results 
achieved so far, the ongoing research and development and the remaining challenges.}

\FullConference{The European Physical Society Conference on High Energy Physics (EPS-HEP2025)\\
7-11 July 2025\\
Marseille, France\\}

\begin{document}
\maketitle

\section{Introduction}
The QCD axion is a cold dark matter (DM) candidate strongly motivated by particle physics. It emerges from the Peccei-Quinn (PQ) symmetry breaking mechanism, proposed in the late 70's to explain the observed absence of CP violation in strong interactions. The only free parameter of this mechanism is the scale of the symmetry breaking, which is orders of magnitude above the electroweak scale. The axion mass ($m_a$) and all its couplings to SM particles are inversely proportional to this scale. The coupling to photons, whose strength is denoted $|g_{a\gamma}|$, is the most widely used experimentally to search for the QCD axion, through its conversion inside a magnetic field to photons. This is how the first haloscopes (resonant cavities searching for the DM axions from the galactic halo) probed a small parameter space of the QCD axion around a few $\mu$eV. The prospects in the next decade to probe a large part of this parameter space (from neV to meV) are promising~\cite{Irastorza:2021tdu}. Notably, the mass range around 100~$\mu$eV is particularly well motivated by the theory to explain the observed DM relic density if the PQ symmetry is broken after the inflation~\cite{OHare:2024nmr}.

\section{The \MADMAX{} experiment}
\MADMAX{}~\cite{MADMAX:2019pub} proposes a novel experimental concept, called dielectric haloscope, to probe this uncharted phase space of the DM QCD axion around 100~$\mu$eV. It uses a mirror and a stack of dielectric disks, called {\it booster}, inside a magnetic field which converts the DM axion field into an electric field, inducing the coherent emission at the different dielectric interfaces of propagating EM waves with a frequency given by the axion mass. If the disks are separated judiciously, these waves could interfere constructively and be resonantly enhanced, boosting the signal significantly (by a boost factor denoted $\beta^2$) over a sizable frequency range~\cite{Millar:2016cjp}. The boost factor quantifies the signal power enhancement with respect to the mirror only set-up, and depends on the dielectric properties and the number of disks. With 80 dielectric discs of $A$=1~m$^2$ surface immersed in a $B_e$=10~T magnetic field, a signal power $P_{sig}$ of the order of $10^{-22}$~W could be achieved for an axion of mass $m_a\sim$100~$\mu$eV, corresponding to $|g_{a\gamma}|\sim2\cdot10^{-14}$~GeV$^{-1}$ for benchmark QCD axion models:
\begin{equation}
\nonumber
    P_{sig} \simeq 10^{-22}~\mathrm{W} 
    \cdot \left( \frac{\beta^2}{5\cdot10^4}    
    \cdot \frac{B_e}{10~\mathrm{T}}
    \cdot \frac{A}{1~\mathrm{m}^2} \right)
    \cdot \left( \frac{|g_{a\gamma}|}{2\cdot10^{-14}~\mathrm{GeV}^{-1}}
    \cdot \frac{100~\mu\mathrm{eV}}{m_a} \right)^2
    \cdot\left( \frac{\rho_a}{0.3~\mathrm{GeV/cm}^3} \right)
\end{equation}
with $\rho_a$ the DM galactic halo density. Such a power could be detected in a few days by state of the art low noise receivers as a narrow peak in frequency above a thermal noise of a few K~\cite{MADMAX:2019pub}. The signal frequency, proportional to $m_a$, ranges between 10 and 100~GHz for 40<$m_a$<400~$\mu$eV. By changing the distance between the disks using piezoelectric motors, the boost factor can be tuned in frequency, enabling to scan the axion mass over a broad range.

The main challenges are the building of a O(10)~T dipole magnet, the calibration of the radio-frequency response of a receiver chain in the $>$10~GHz regime operated at cold, and the mechanical control at the O($\mu$m) level of movable disks of O(1)~m diameter at a few K and in a high magnetic field. The \MADMAX{} collaboration thus started in 2020 a prototyping phase, to validate the dielectric haloscope concept and gradually build the final booster. Small-scale prototype boosters composed of sapphire disks were operated at CERN in a dipole magnet of 1.6~T, enabling first dark matter searches in an uncharted phase space.

\section{First dark matter searches with \MADMAX{} prototypes}
\begin{figure}[htbp]
    \centering
    \includegraphics[width=0.99\textwidth]{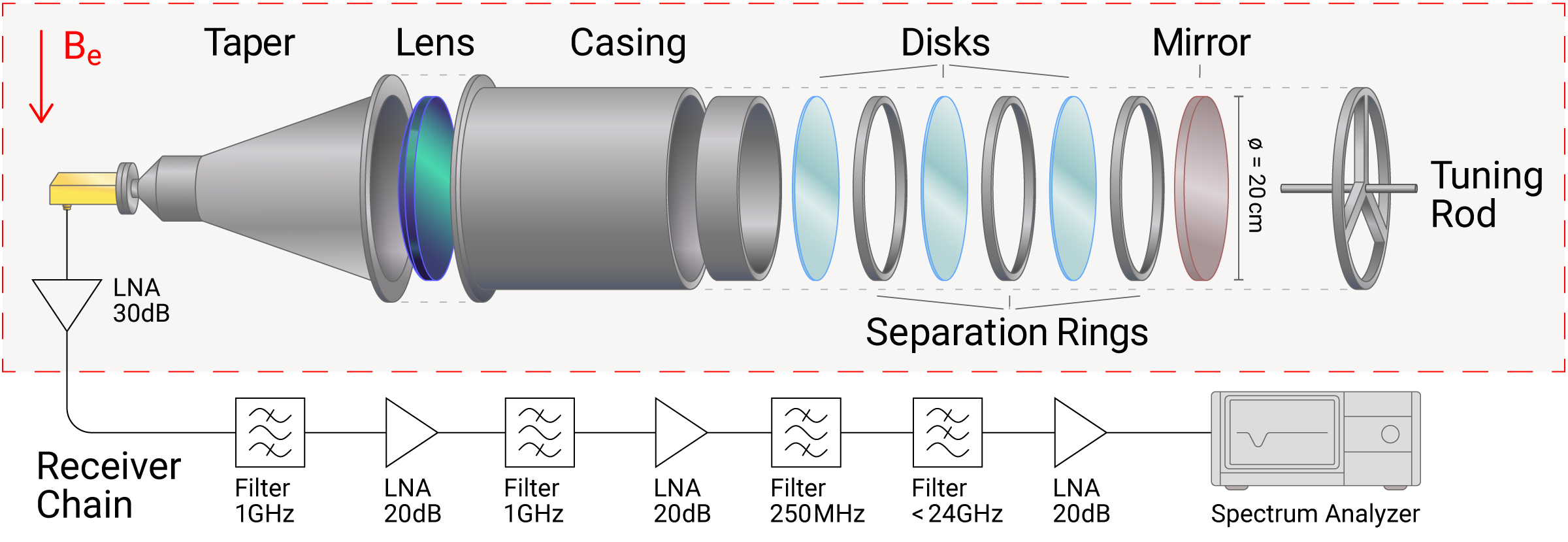}
    \caption{Exploded schematic view of the \MADMAX{} booster prototype and the receiver chain used for the first search for axion dark matter~\cite{MADMAX:2024sxs}.}
    \label{fig:CB200}
\end{figure}

The first ever search for DM axion with a dielectric haloscope was performed at room temperature with the booster prototype shown in Figure~\ref{fig:CB200}. It is made of a mirror and three sapphire disks of 20~cm diameter. The disks were not moving dynamically, their distance was fixed with separation rings. Two different sets of separation rings were used, allowing to tune the booster peak frequency around 18.55 and 19.21~GHz, corresponding to axion masses of 76.72 and 79.45~$\mu$eV. Moreover, a tuning rod allowed to move the mirror by O(10)~$\mu$m inducing frequency variations of O(10)~MHz, resulting in five different configurations. 15 days of data were taken in 2024 inside a magnetic field between 1 and 1.6~T. 

\begin{figure}[htbp]
    \centering
    \includegraphics[width=0.54\textwidth]{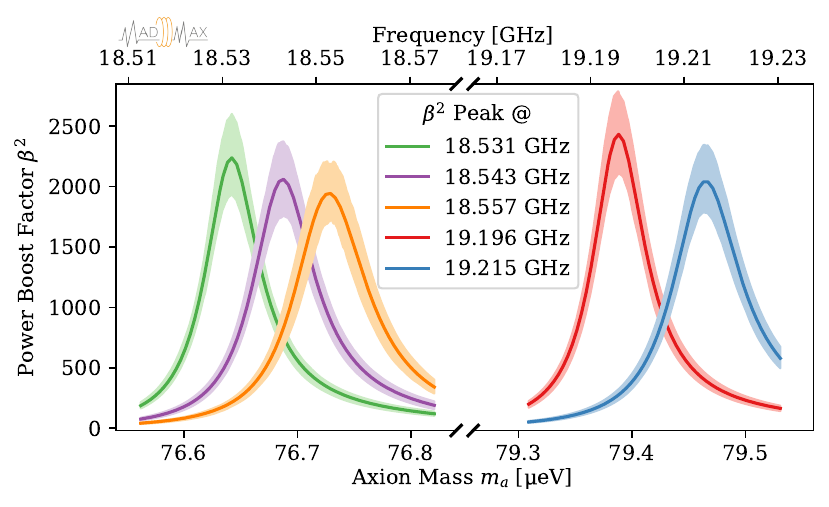}
    \includegraphics[width=0.45\textwidth]{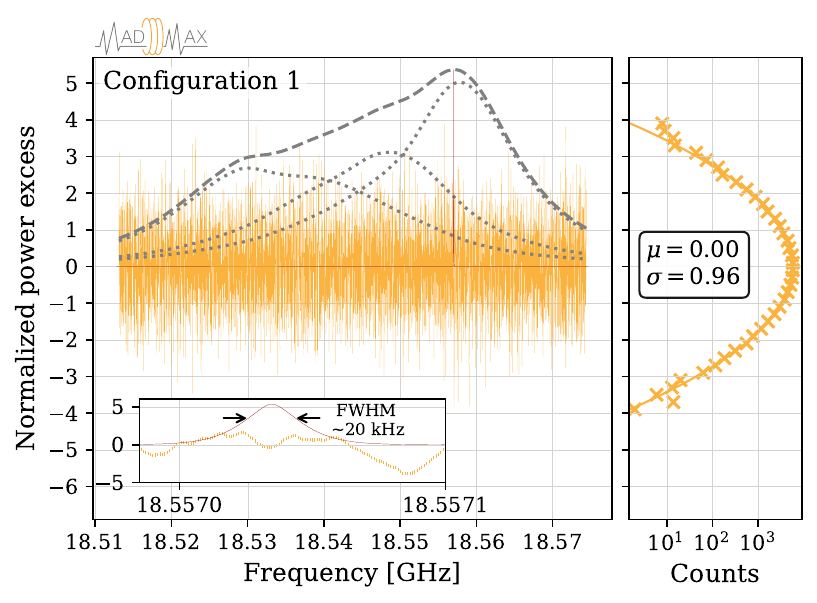}
    \includegraphics[width=0.7\textwidth]{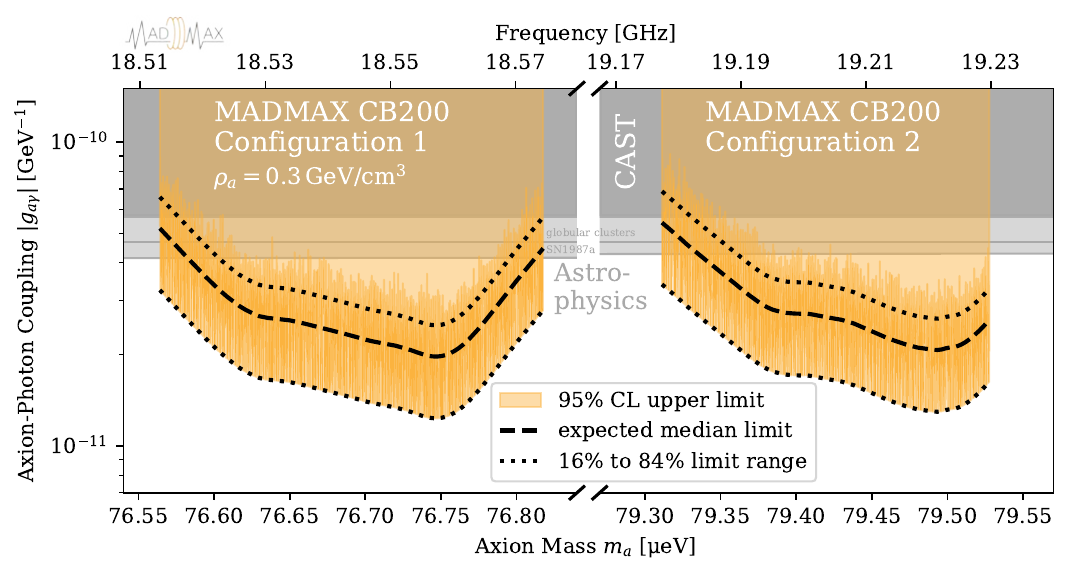}
    \caption{Results of the first DM axion search with a \MADMAX{} prototype~\cite{MADMAX:2024sxs}. Top left: boost factor distributions and associated uncertainties for the five configurations. Top right: measured power excess with respect to the thermal noise (orange) around 18.55~GHz, with its projection in the right panel. A reference axion signal with $m_a$=76.75~$\mu$eV and $|g_{a \gamma}| = 3.5\times 10^{-11}\,$GeV$^{-1}$ is shown in red, with a zoom in the inset. Bottom: 95\% CL exclusion limits in the $m_a$--$|g_{a \gamma}|$ plane.} 
    \label{fig:axion}
\end{figure}

The disks were encapsulated in a cylindrical waveguide (called {\it Casing} in Figure~\ref{fig:CB200}), so that only the fundamental transverse electric mode could propagate inside the booster. This very much simplifies the modeling of the electromagnetic response of the booster and allows to determine the boost factor curves from dedicated measurements of the reflectivity and of the noise of the system. As shown in Figure~\ref{fig:axion} (top left), the $\beta^2$ distributions peak around~2000 for the five configurations, with 15\% systematic uncertainties. 

The data analysis did not reveal any significant excess above the thermal noise, as illustrated in Figure~\ref{fig:axion} (top right) for the low frequency configurations. Upper limits on $|g_{a\gamma}|$ are thus derived in two axion mass ranges of size 0.2~$\mu$eV (Figure~\ref{fig:axion} bottom)~\cite{MADMAX:2024sxs}. They are exceeding the existing limits in these mass ranges from the CAST helioscope and from astrophysical constraints, despite the modest size of the system. This confirms the substantial potential of the dielectric haloscope concept. 

A further increase in sensitivity can be reached by cooling down the system, reducing the thermal noise. A similar prototype has been operated in the dipole magnet at CERN and in a novel type of cryostat made of non-magnetic glass-fiber~\cite{Kreikemeyer-Lorenzo:2024efy}, allowing to reach temperatures below 10~K for more than a day. Data analysis is underway.\\

\clearpage

\begin{figure}[htbp]
    \centering
    \includegraphics[width=0.43\textwidth]{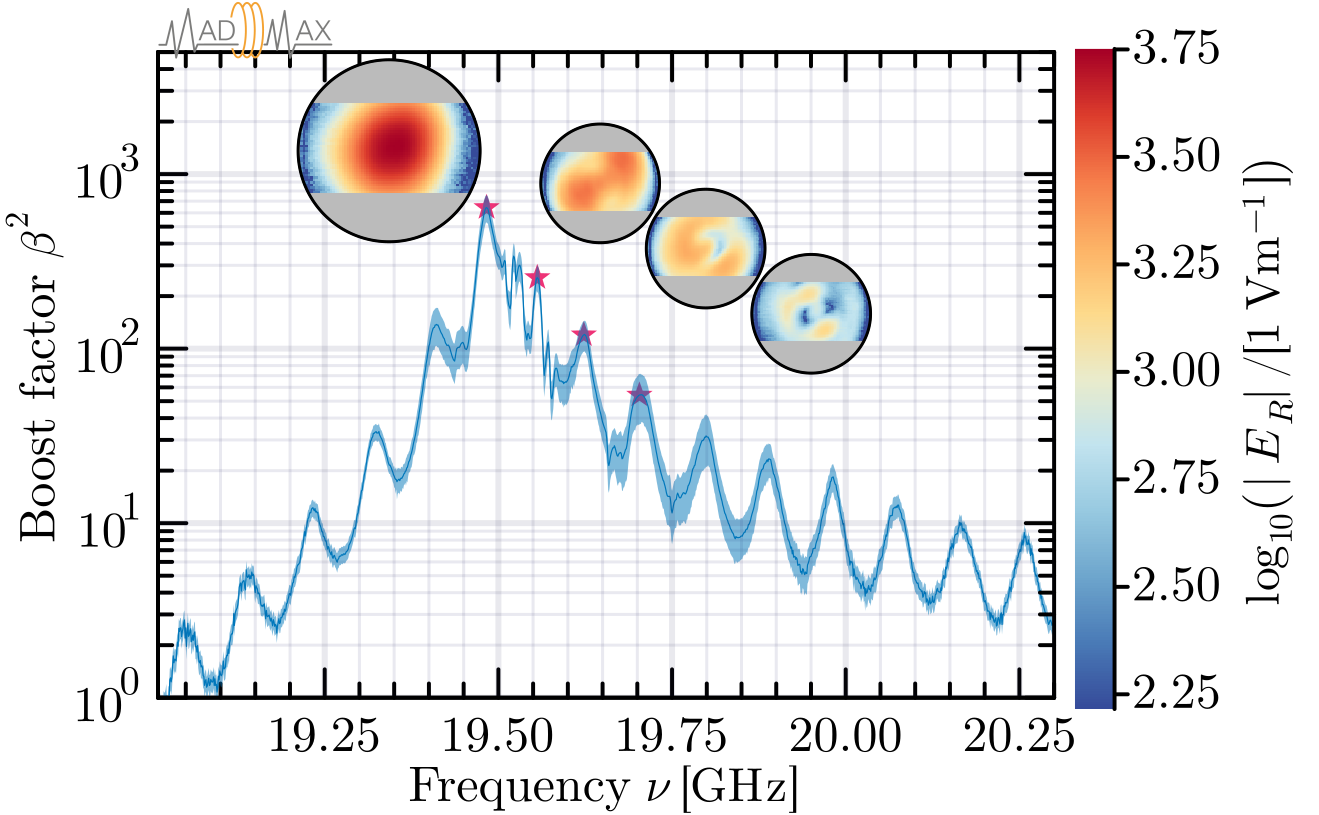}
    \includegraphics[width=0.55\textwidth]{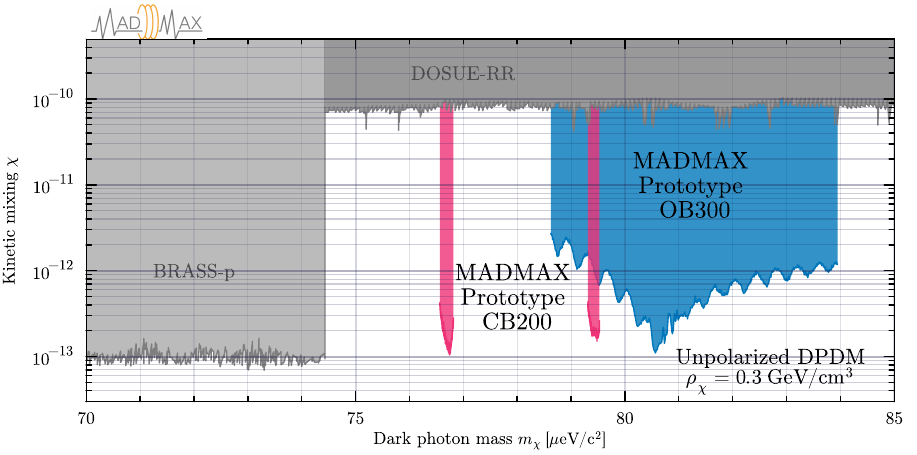}
    \caption{Results of the first DM dark photon search with a \MADMAX{} prototype~\cite{MADMAX:2024jnp}. Left: measured boost factor distribution and associated uncertainties. Right: 95\% CL exclusion limits in the $m_\chi$--$\chi$ plane, where $\chi$ is the kinetic mixing angle between photons and dark photons.} 
    \label{fig:dp}
\end{figure}

The next step was to use a prototype with three bigger (30~cm diameter) fixed disks without a waveguide around them, which will allow in the future to move them dynamically (see later). In this case, a small-sized perturbing dielectric object can easily be inserted between the disks, allowing to measure in situ the boost factor from the induced change in the booster reflection coefficient~\cite{Egge:2022gfp,Egge:2023cos}. The measured boost factor distribution is shown in Figure~\ref{fig:dp} (left)~\cite{MADMAX:2024jnp}. The oscillations are due to higher-order transverse modes (with respect to the fundamental gaussian mode) resonating between the antenna and the booster, as illustrated by the insets showing the reflection-induced transverse electric field between the mirror and the first disk at the indicated frequencies (stars). The $\beta^2$ distribution peaks around 640 with a measurement uncertainty of 15\%, and it is above~1 over a broad frequency range of 1.3~GHz. A search for DM dark photons was conducted with 12 days of data taken with this prototype operated at room temperature without magnetic field in a shielded laboratory. No significant excess above the thermal noise was revealed, allowing to put upper limits on the kinetic mixing angle between photons and dark photons in the mass range from 78.6 to 84~$\mu$eV (Figure~\ref{fig:dp} right)~\cite{MADMAX:2024sxs}. They are exceeding the existing limits in this mass range by up to 3 orders of magnitude, despite the modest size of the system. The DM axion limits from Figure~\ref{fig:axion} have also been reinterpreted in terms of dark photon mixing, as shown in pink.   

\section{Towards the final \MADMAX{} experiment}
To scan over a broad axion mass range, the disks of the booster must be movable, with their position controlled at the O($\mu$m) level. This has been successfully demonstrated with a set-up made of one disk moved by three piezoelectric motors~\cite{Garutti:2023stk} whose position was controlled by laser interferometers. Both inside a magnetic field of 1.6~T and at cryogenic temperatures down to 35~K, the measurements of the velocity and of the positioning accuracy of the disk were found to meet the \MADMAX{} requirements~\cite{MADMAX:2024pil}, validating the mechanics of the booster.

\begin{figure}[htbp]
    \centering
    \includegraphics[width=0.43\textwidth]{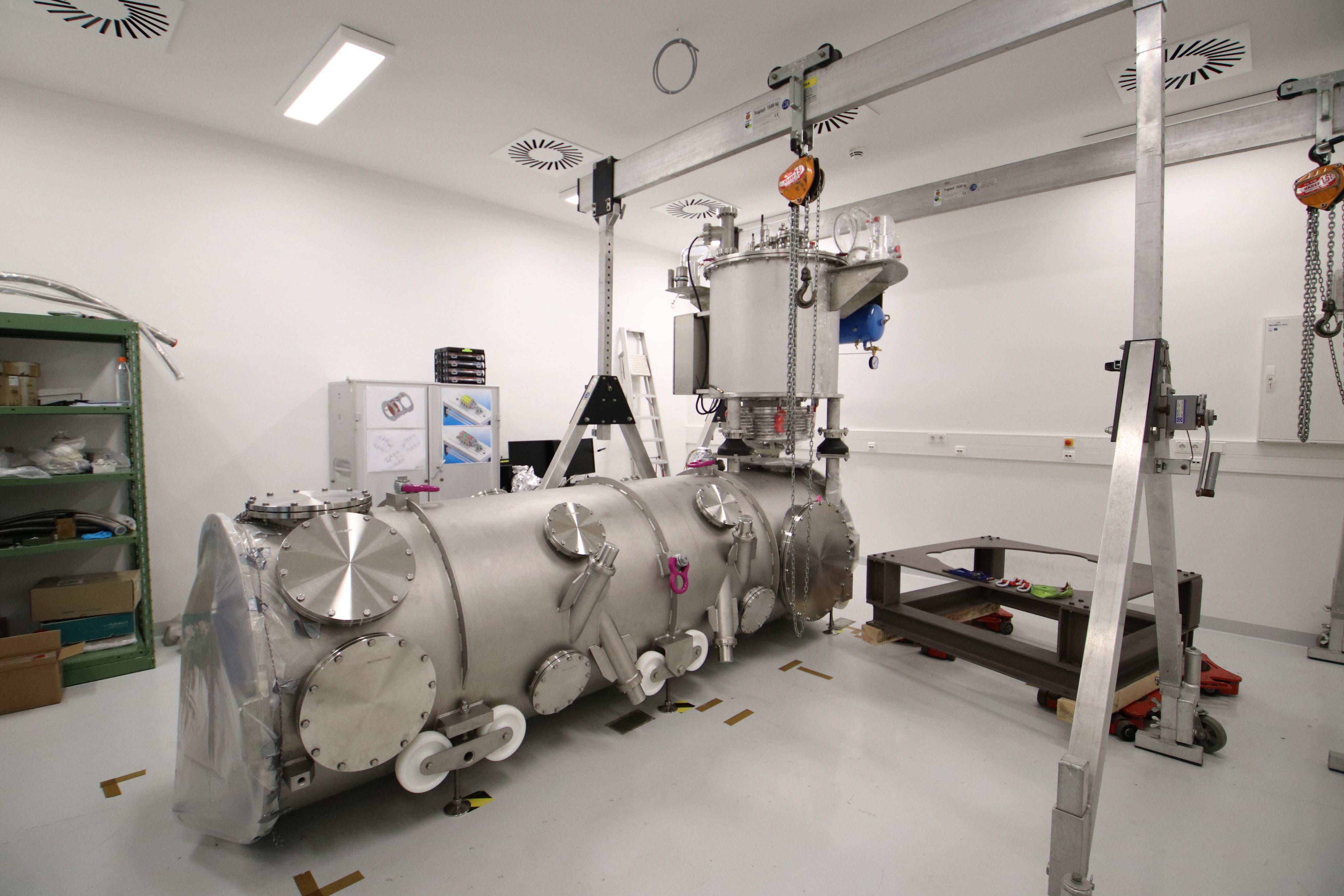}  
    \includegraphics[width=0.43\textwidth]{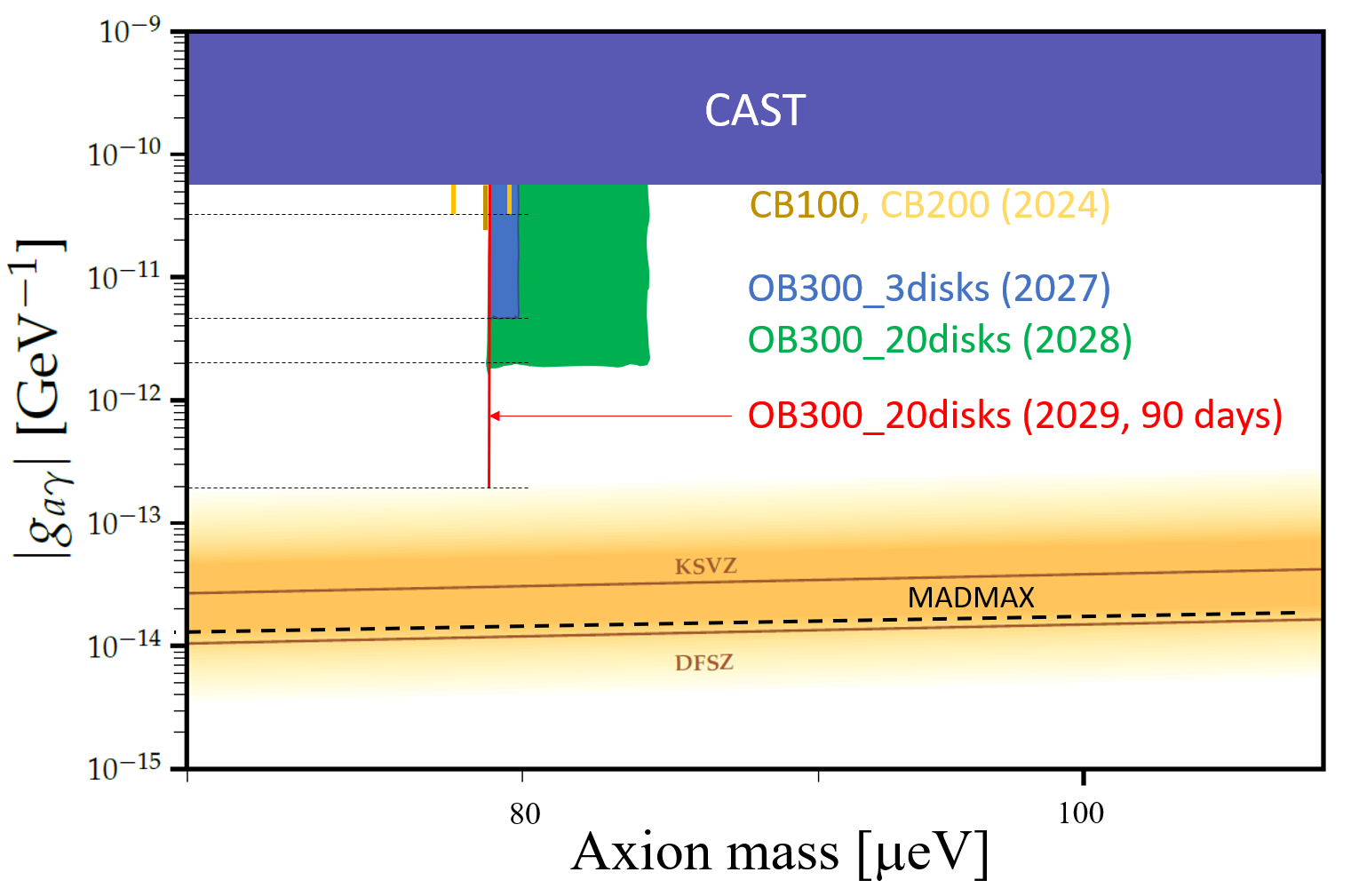}
    \caption{Left: picture of the cryostat to operate the final \MADMAX{} prototype. Right: physics reach in the $m_a$--$|g_{a \gamma}|$ plane of this prototype (called OB300) in the near future, as compared to current results.} 
    \label{fig:next}
\end{figure}

The ultimate step of the prototyping phase is to gather all the developments reported so far to build a booster made of 3 to 20 disks of 30~cm diameters that can be moved precisely, and operate it at cryogenic temperatures in the CERN dipole magnet. The stainless steel cryostat to cool down the system down to 4~K has been designed and constructed (Figure~\ref{fig:next} left). This will allow to perform axion searches during the LHC shutdown in 2027-2029, with three months runs per year, gaining in sensitivity by up to two orders of magnitude compared to current results and scanning the axion mass over a limited but sizable range, as illustrated in Figure~\ref{fig:next} (right).

By validating the power and versatility of the dielectric haloscope concept, this prototyping phase will pave the road towards the design of the final \MADMAX{} detector. The later will be operated at DESY in order to search for the DM QCD axion in a large mass range around 100~$\mu$eV, favored by post-inflationary theories. 

\end{document}